\documentclass[twocolumn,showpacs,preprintnumbers,amsmath,amssymb,prl]{revtex4-1}

\usepackage{graphicx}
\usepackage{dcolumn}
\usepackage{bm}
\providecommand{\boldsymbol}[1]{\mbox{\boldmath $#1$}}
\newcommand{\br}{{\boldsymbol r}}
\newcommand{\brp}{{\boldsymbol r}^\prime}

\begin{document}

\title{Direct Statistical Simulation of Out-of-Equilibrium Jets}

\author{S. M. Tobias}
\email{smt@maths.leeds.ac.uk}
\affiliation{Department of Applied Mathematics, University of Leeds, Leeds, LS2
9JT, U.K.}

\author{J. B. Marston}
\email{marston@brown.edu}
\affiliation{Department of Physics, Box 1843, Brown University,
  Providence, RI 02912-1843, USA}

\date{\today}

\begin{abstract}
We present Direct Statistical Simulation (DSS) of jet formation on a $\beta$-plane, 
solving for the statistics of a
fluid flow via an expansion in cumulants.   Here we compare an expansion truncated at
second order (CE2) to statistics accumulated by direct numerical simulations (DNS).  We show that, for
jets near equilibrium, CE2 is capable of
reproducing the jet structure (although some differences remain in the
second cumulant). However as the degree of departure from equilibrium
is increased (as measured by the zonostrophy parameter) the jets
meander more and CE2 becomes less accurate. We
discuss a possible remedy by inclusion of higher cumulants.

\end{abstract}


\maketitle

Jets are relatively narrow bands of fast-flowing fluid moving 
coherently in one direction.  They are ubiquitous in nature, found in
Earth's oceans and atmosphere, the outer layers of gas giant
planets, the interior of stars, and laboratory experiments with
fluids and plasmas \cite{diamondetal05}.  Jets play an important role for
these fluids, and it is therefore important to understand the
mechanism(s) that govern their formation. Sometimes
jets are driven by energy input at small spatial scales; the question
then is how this energy is transferred into large scale coherent
motion.   Two competing mechanisms have been proposed, both of which rely on the interaction of turbulence and rotation. The first
involves the scale-by-scale transfer of energy known as the inverse
cascade \cite{valmal93}.  Large-scale vortices are known to be generated by this
mechanism.  The other mechanism relies upon direct transfer of energy
to the largest scales.  It is hard to disentangle these two mechanisms
in experiments and in simulations.  Direct calculation of statistics and quasilinear direct numerical simulation (DNS) calculations have demonstrated that jets can be
formed by the direct mechanism, not relying on an inverse cascade \cite{OGorman:2007wn,Farrell:2007fq,Tobiasetal11,sriyoung12} and see also \cite{huangrobinson98}.   

Non-equilibrium statistical mechanics can be used to understand universal aspects of fluids.  Isotropic, homogeneous, turbulence is at present beyond the reach of a complete
statistical theory.  By contrast, inhomogeneous flows such as jets may be accessible to Direct Statistical Simulation (DSS), that is, methods
solving directly for the statistics of the flow \cite{lorenz67}.  DSS offers the possibility of a deeper understanding fluid dynamics, as well as a practical speed-up in obtaining statistics \cite{Tobiasetal11}.  
In the limit of small driving  and dissipation, equilibrium statistical mechanics is a powerful tool for understanding quasi two-dimensional
flows (for a review see Ref. \cite{Bouchet:2012ea}).  Here and below the word ``equilibrium'' refers to the limiting case for which the rates of forcing and dissipation go to zero.
Away from equilibrium, 
Stochastic Structural Stability Theory (SSST) \cite{Farrell:2007fq,farrellioa08,Constantinou:2012ui} is one approach that has 
been explored to understand the formation and maintenance of jets.   Here we instead investigate systematic expansion 
in equal-time, but spatially nonlocal, cumulants of the flow.  When truncated at second order, the cumulant expansion (denoted CE2)
is closely related to SSST \cite{sriyoung12} but it is only the starting point for a perturbative expansion in higher cumulants.  

This paper examines the accuracy of DSS at CE2
as a representation of the statistics of turbulent flows driven  away from
equilibrium.  CE2 includes the interaction of
mean flows with eddies to drive eddies and that of eddies
with eddies to drive mean flows, but removes 
the interaction of eddies with eddies in the evolution equation for
the eddies \cite{marconschn08}; an interaction that has been termed the ``EENL'' (eddy-eddy
nonlinearity) by Srinivasan and Young \cite{sriyoung12}. Here eddies are formally the fluctuations about the zonal mean flow. It has been argued \cite{Bouchet:2012ea,Bouchet2012pc}
that CE2 is an exact representation in the quasi-equilibrium limit, 
but the domain of validity of such a truncation  remains largely untested.  We conduct numerical experiments to investigate the accuracy of 
DSS at CE2 for systems removed from
quasi-equilibrium by considering a model problem of the driving of
jets by small-scale forcing 
on a $\beta$-plane. This system has been
studied extensively within the framework of DNS in both the fully
nonlinear and quasi-linear regimes
\cite{scottdri12,sriyoung12}. Although this model is the 
simplest that includes all the requisite features for our purposes,
i.e. anisotropy, non-trivial long-range correlations and mean flows,
we note that it is a rigorous test of
statistical methods in that it is stochastically  driven and translationally invariant in two
directions, with only the emergence of jets spontaneously breaking the latitudinal
symmetry \cite{sriyoung12} and leading to inhomogeneity. We return to this in the discussion
at the end of the paper.

The $\beta$-plane we use is periodic in both $x$ (longitude) and $y$ (latitude), 
with the domain of size $2 \pi \times 2 \pi$.  
The motion of the incompressible fluid is damped by a single friction
$\kappa$ and  by
small-scale dissipation that absorbs structures at the finest scales. 
(Some models examined in Ref. \cite{Constantinou:2012ui} have friction damping the fluctuations 10
times greater than that slowing the zonal mean flow.)
The fluid is driven by random (stochastic) forcing $\eta$.   This type of
stochastic forcing is widely used as a model of small-scale
processes that inject energy into the fluid, with the small and fast
scales acting as a random influence on the large and slow
scales \cite{Chandrasekhar:1943eb,Hasselmann:1976ua,vandenEijnden:1998gu}.
The time-evolution of the relative vorticity $\zeta \equiv \hat{z}
\cdot (\vec{\nabla} \times \vec{u})$ is given  by for example
\cite{galsukdik08}:  
\begin{eqnarray}
\partial_t \zeta + J(\psi,\zeta) + \beta \partial_x \psi &=& -\kappa \zeta+\nu \nabla^2 \zeta+\eta, \label{z_eqn1}\\
\zeta &=& \nabla^2 \psi,
\label{z_eqn2}
\end{eqnarray}
where $J(a,b) = \partial_x a~ \partial_y b - \partial_y a~ \partial_x b$. Here $\psi$ is the streamfunction
and the fluid velocity
$\vec{u} = \left(u,v\right)= \left(-\partial_y \psi,~ \partial_x \psi \right)$; 
we have set the deformation radius of the flow to be infinite. The forcing is  random with a short (but
non-zero) renewal time ($0.1 \le r_t \le 1$) and concentrated in the spectral band of
wavenumbers $k_{min} \le |k_x|, |k_y| \le k_{max}$ (for these runs $k_{min}=7 $, $k_{max}=10$). 
The amplitude of the forcing is chosen from a Gaussian distribution with standard deviation $a_\eta$. 
This is a popular choice of forcing; a detailed discussion of the role of
forcing in DNS of such problems is given in Ref. \cite{scottdri12}. 

Rhines \cite{rhin75}, who investigated the unforced system, demonstrated how
correlations between nonlinear Rossby waves could lead to the
generation of zonal flows and identified the scale at which zonal
flows become important in mediating the dynamics of these waves (see
e.g. Ref. \cite{vallis06}). This ``Rhines scale'' is given by
$L_R=(2U/\beta)^{\frac{1}{2}}$, where $U$ is the rms velocity of the flow,  and occurs when the second and third terms of
equation~(\ref{z_eqn1}) are comparable  (and are comparable with the frictional term \cite{sdg:2007}). There has been much research into the
importance of this lengthscale for the ultimate latitudinal scale of
jets (see e.g. Ref. \cite{dritmcintyre08} and the references therein), but it
is also becoming clear that the dynamics of zonation is also
controlled by another length scale $L_\varepsilon$
\cite{maltrudvallis91}, which measures the intensity of the forcing 
relative to the background potential vorticity gradient. For the
simple $\beta$-plane model  $L_\varepsilon=
0.5(\varepsilon/\beta^3)^{\frac{1}{5}}$ where $\varepsilon$ is the energy input
rate of the stochastic forcing $\eta$. 

The ratio of these two length scales has been proposed, for
  models with small-scale forcing, to play a critical role in
determining the strength and stability of jets
\cite{galsukdik08,galetal2010},  
for cases where the same damping is applied to the mean flows and the
turbulent fluctuations.  
This local measure, termed the zonostrophy index, is given by $R_\beta
\equiv L_R / L_\varepsilon = U^{1/2} \beta^{1/10}/2^{1/2}\varepsilon^{1/5}$. In general, if the zonostrophy
index is large then strong stable jets are
found, whilst for small $R_\beta$ the jets are weaker, meander more
and no staircase is formed \cite{scottdri12}. The zonostrophy
index is therefore a measure of how far the system is driven out of
equilibrium. Note that $R_\beta$ can also be written (on balancing the energy input with the dissipation via friction $\varepsilon \sim \kappa U^2$) in terms
of the ratio of  an advective time on the
Rhines scale to a dissipative timescale   ($F_\beta=\kappa L_R /U$)
i.e.\ $R_\beta=F_\beta^{-1/5}$. 
 Hence the quasi-equilibrium limit is given by $R_\beta
\rightarrow \infty$. Recent estimates have put $R_\beta$ between  $5$ and $6$ for
flows on the surface of Jupiter \cite{Galperin2012pc}, whilst $R_\beta
\sim 2$ for oceanic  
jets \cite{galsukdik08}. We note that the zonostrophy index might not be the only parameter controlling the dynamics of the jets. It has been shown  that if the forcing lengthscale remains important then the dynamics is controlled by two non-dimensional parameters separately\cite{sriyoung12}, and there is a regime given by a chain inequality for which $R_\beta$ is the only important non-dimensional parameter\cite{galsukdik08}. Nonetheless, even in this regime $R_\beta$ does give {\it a} measure of lack of equilibrium. 
\begin{figure}

\centerline{\includegraphics[width=2in]{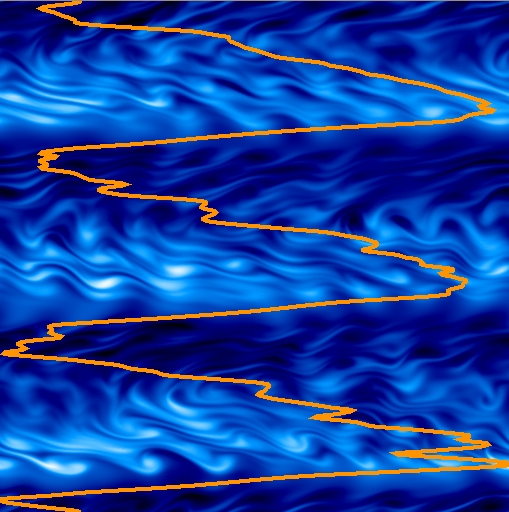}}
\caption{\label{figure1} Snapshot density plot of vorticity together with zonal mean vorticity profile of jets found by DNS. 
The parameters are  $\kappa=10^{-3}$, $\nu=10^{-4}$, $\beta=16$.  For these parameters $R_\beta = 2.12$.}
\end{figure}

\begin{figure}
\centerline{\includegraphics[width=3.0in]{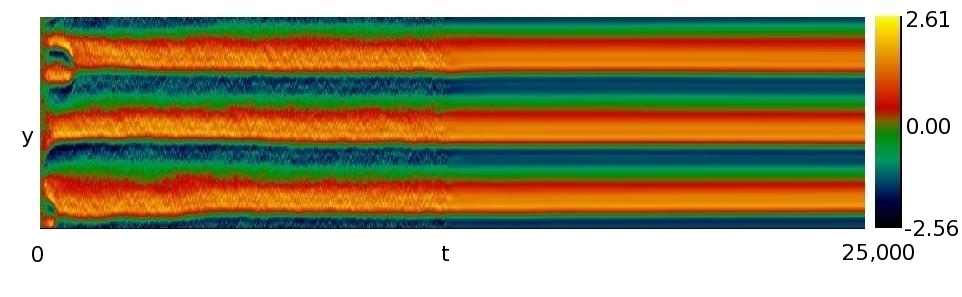}}
\vskip 0.1cm
\centerline{\includegraphics[width=3.0in]{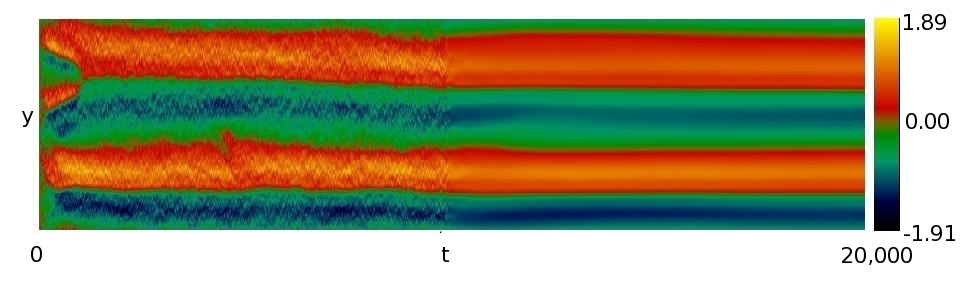}}
\vskip 0.1cm
\centerline{\includegraphics[width=3.0in]{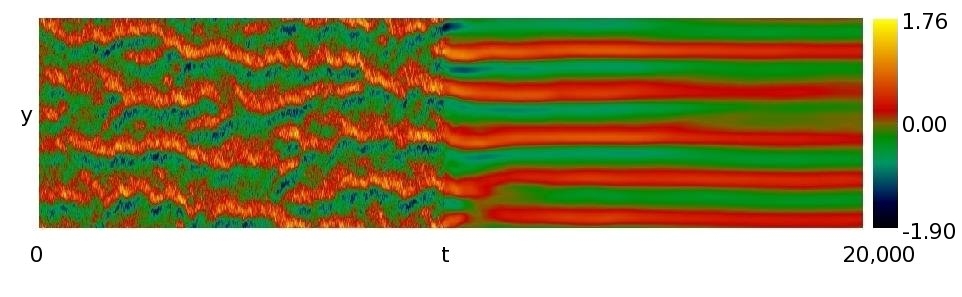}}
\caption{\label{figure2} Hovm\"oller diagrams of zonal mean relative vorticity versus time from DNS simulations. (a) Parameters as for Fig.~\ref{figure1} with $R_\beta = 2.12$; (b) Parameters as for (a) but $\beta=8$ and $R_\beta = 1.98$; here $L_R$ is increased by a factor $3/2$ from that in (a); (c) Parameters as for (a) but $\kappa=10^{-2}$ and $R_\beta = 1.24$; here $L_R$ is decreased by a factor $3/5$ from that in (a).  A running time-average commences at the midpoint of each diagram.}
\end{figure}

DNS is performed using a pseudo-spectral scheme optimised for parallel machines \cite{tobcatt08}. For these simulations
we utilise resolutions of up to $2048^2$. The forcing is applied at
moderate scale  (with $r_t a_{\eta}^2=0.01$ for all calculations) and the
system is evolved from rest until a statistically steady
state is reached. Fig.~\ref{figure1} shows a snapshot of the vorticity and zonally
averaged vorticity for a state with 3 zonal jets. For this calculation
$R_\beta =2.12$ and 
the jet is well removed from the quasi-equilibrium limit. We note that
this limit is difficult to  
simulate in DNS, requiring long integrations. Nonetheless
the Hovm\"oller diagram in Fig.~\ref{figure2}(a) of the
$(t, y)$ dependence of the mean flow together with a running time
average calculated from the midpoint of the calculation shows that the
zonal flows do not meander too much in space and well-defined averages
can be obtained --- though we note
that lengthy integrations of the dynamics are required for
meaningful flow statistics.  Fig.~\ref{figure2}(b) shows the
corresponding diagram when $R_\beta$ has been reduced to $1.98$, which
is achieved here by lowering $\beta$. For this case, even further from
quasi-equilibrium,  the jets are still relatively steady, but the
Rhines scale has changed sufficiently that now only two jets fit in
the domain. This is consistent with the values of $L_R$ given in the figure caption. For Fig.~\ref{figure2}(c) $R_\beta$
has been reduced to $R_\beta=1.24$, achieved by increasing the
friction. Here the jets meander significantly and merge showing a
large degree of spatiotemporal variation. In this respect they have
the characteristics of observations and simulations of oceanic jets
\cite{maximenkoetal05,nadiga06}. At different times there appears either four
or five jets, but on average there are five jets. Because of the
temporal variability, the time average of the jet velocity is much
smaller than the instantaneous jet speeds. 

Expansion of equal-time cumulants at order CE2 is straightforward for
Eq. \ref{z_eqn1}.   
Let $\br = (x,y)$ and $\brp = (x^\prime,y^\prime)$ and adopt a Reynolds decomposition by setting
$\zeta(\br) = \langle \zeta(\br) \rangle + \zeta^\prime(\br)$,  where
the angle brackets imply either an ensemble average or an average over
longitude ($x$). We define the first cumulants as
$c_\zeta=\langle \zeta(\br) \rangle=c_\zeta(y)$ and 
$c_\psi(\br) = \langle \psi(\br)\rangle = c_\psi(y)$ where the relationship between these is given by 
$c_\zeta ={\partial^2_{yy}}  c_\psi$.
We may then define the second cumulants as follows:
$
c_{\zeta \zeta}(\br,\brp) =  \langle \zeta^\prime(\br) \zeta^\prime(\brp)\rangle
$
and note that for this system $c_{\zeta \zeta}$ depends on the two local latitudes and the difference between the longitudes $\xi=x-x^\prime$, i.e.\
$
c_{\zeta \zeta}(\br,\brp) = c_{\zeta \zeta}(y,y^\prime,\xi)
$ \cite{marconschn08}.
Corresponding definitions arise for the derived second cumulants $c_{\psi \zeta}$ and $c_{\zeta \psi}$, i.e.\
$c_{\psi \zeta}(\br,\brp) =  \langle \psi^\prime(\br) \zeta^\prime(\brp)\rangle=c_{\psi \zeta}(y,y^\prime,\xi)
$
and similarly for $c_{\zeta \psi}(y,y^\prime,\xi)$.  With these definitions the equations for cumulant hierarchy, truncated at second order, are
 \begin{eqnarray}
{\partial_t}  c_{\zeta} &=&
 -\left({\partial_y}+ {\partial_{y^\prime}}\right){\partial_\xi} c_{\psi \zeta}|_{y \rightarrow y^\prime}^{\xi \rightarrow 0}  - \kappa c_\zeta+
\nu {\partial^2_{yy}} c_\zeta . 
\label{czeqn}\\
{\partial_t}  c_{\zeta \zeta} &=& 
{\partial_y} c_\psi {\partial_\xi} c_{\zeta \zeta} 
- {\partial_y} (c_\zeta(y)-\beta y) {\partial_\xi} c_{\psi \zeta} \nonumber \\
&-&{\partial_{y^\prime}} c_\psi {\partial_\xi} c_{\zeta \zeta} 
+{\partial_{y^\prime}} (c_\zeta(y^\prime)-\beta y^\prime){\partial_\xi} c_{\zeta \psi} \nonumber \\
&+&\nu \left( \nabla^2+\nabla'^2\right) c_{\zeta \zeta} -2 \kappa c_{\zeta \zeta}+ \Gamma.
 \label{czzeqn}
\end{eqnarray}
Here $\Gamma$ is the covariance matrix of the stochastic forcing that enters into equation~\ref{czzeqn}  as a deterministic source term localised at the same wavenumbers as for the DNS \cite{Tobiasetal11} and with an amplitude $a_\Gamma$ that is given $a_\Gamma= r_t a_\eta^2$. Eqs. \ref{czeqn} and \ref{czzeqn} constitute a realizable closure and in the absence of damping and forcing, conserve linear momentum, energy, and enstrophy.
Eqs. \ref{czeqn} and \ref{czzeqn} are integrated forward in time using a
pseudo-spectral  integrating factor/Adams
Bashforth numerical scheme. The integrations were performed at a typical resolution of 
$16 \times 128$.  Restricting $|k_x| < 16$ does not amount to a further approximation beyond CE2, because, for this problem, at level CE2 only modes with zonal wavenumbers less than 
those of the stochastic forcing are excited. 

\begin{figure}
\centerline{\includegraphics[width=3.0in]{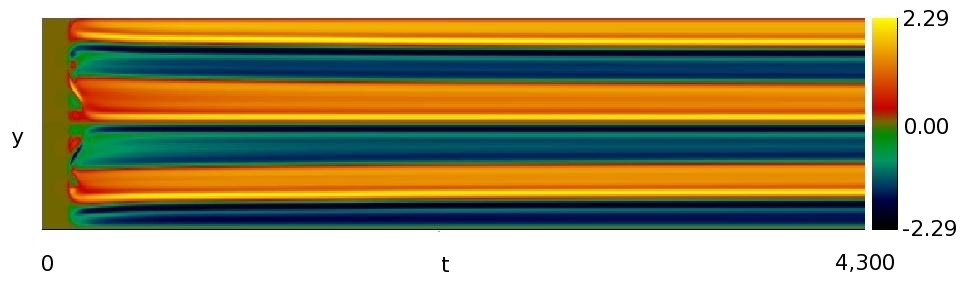}}
\centerline{\includegraphics[width=3.0in]{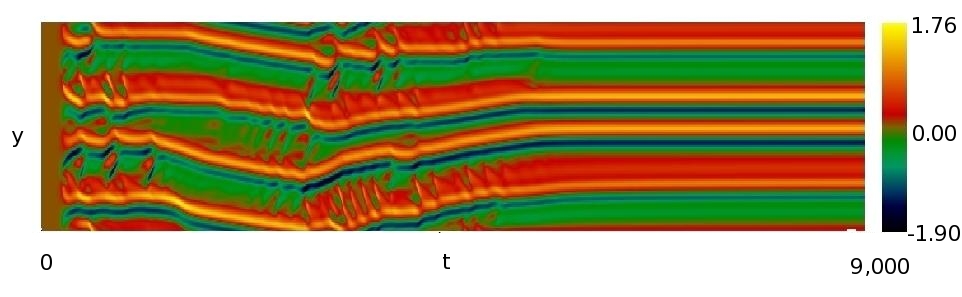}}
\caption{\label{figure3} Hovm\"oller diagrams of relative vorticity from CE2. (a) Parameters as for Fig.~\ref{figure2}(a).  (b) Parameters as for Fig.~\ref{figure2}(c).}
\end{figure}

\begin{figure}
\centerline{ \includegraphics[width=3.8in]{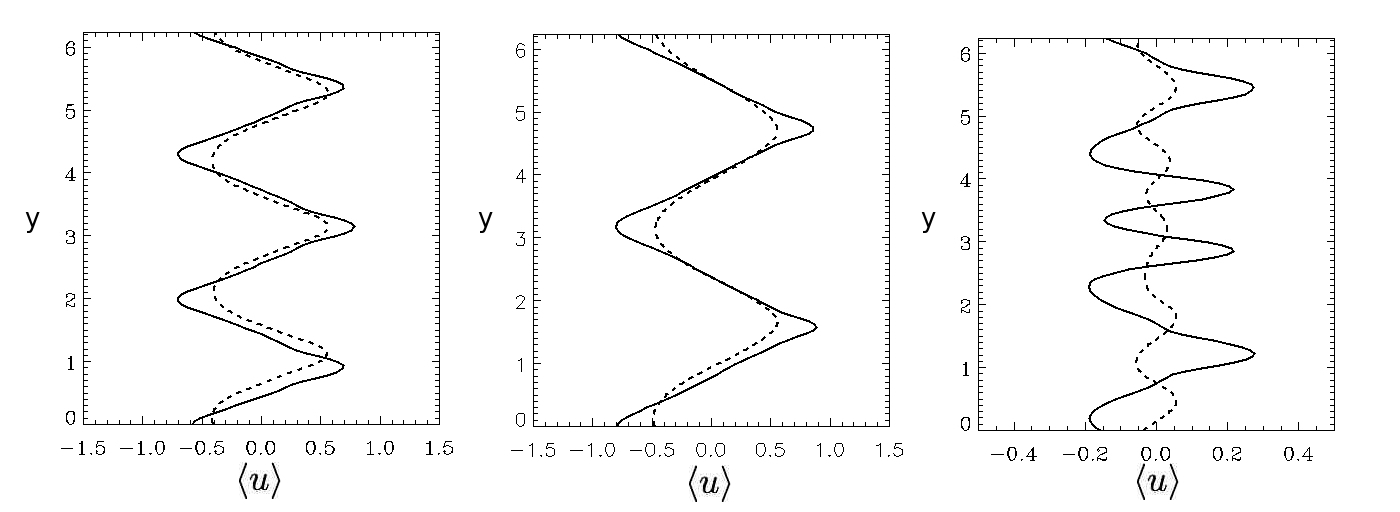}}
\caption{\label{figure4} Comparison of mean zonal velocity from DNS (dashed lines) and CE2 (solid lines) for parameters as in Figs.~\ref{figure2}(a), (b), and (c)
for which $R_\beta = 2.12$, $1.98$, and $1.24$.}
\end{figure}

\begin{figure}
\centerline{\includegraphics[width=3.2in]{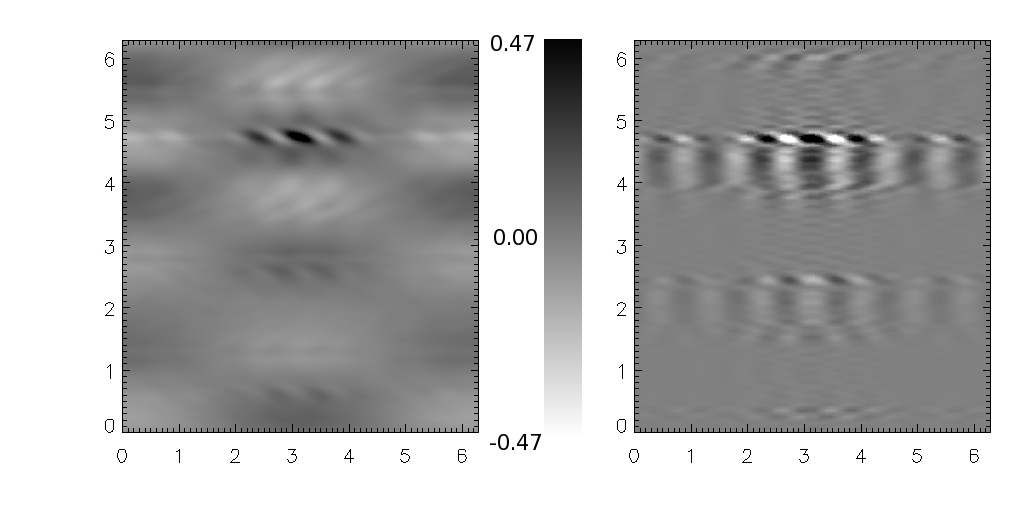}}
\caption{\label{figure5} Second cumulant $c_{\zeta \zeta}$ as calculated from DNS (left) with $R_\beta = 2.12$ and the corresponding CE2 solution (right). 
Cross correlation with respect to a test point located at $(\pi,4.7)$.}
\end{figure}

Recall that for $R_\beta$ large the system is in quasi-equilibrium, dominated by strong jets, 
and CE2 should provide an accurate representation of the statistics of the fluid flows.
A typical evolution of the cumulant system is shown
in Fig.~\ref{figure3}(a). After some
initial transients where jets are driven with a relatively small
latitudinal extent, broader jets emerge via a series of jet
mergings. Similar jet-merging behaviour has been observed both in DNS
and in the
strong jet simulations of SSST \cite{farrellioa08}, and also in the weakly nonlinear description of zonal jets \cite{manfroiyoung99}. The system eventually reaches a
statistically steady state, represented by a simple fixed point of the
cumulant equations. The calculations
were repeated at a range of $R_\beta$ and compared with the (zonal and time averages
of the) DNS solutions described earlier.  Fig.~\ref{figure4} shows
comparisons of the zonal velocity in the jet from DNS averaged over both $x$ and
time with that achieved from DSS at CE2 for $R_\beta = 2.12$ and
$R_\beta=1.98$. The agreement in the first cumulant
at these levels of disequilibrium is good; CE2 reproduces both
the correct number of jets and their strength; although CE2 slightly overestimates the average jet strength 
--- a characteristic in common with quasi-linear DNS of jets \cite{sriyoung12}.  However close
examination of the second cumulant reveals that CE2 struggles to
reproduce the cross-correlation patterns (or teleconnections) from DNS
for these parameters. The left panel of Fig.~\ref{figure5} shows the second
cumulant as accumulated from the DNS solution of Fig.~\ref{figure1}. The
figure shows the cross-correlation of the vorticity statistics with respect to a test point.  The
second cumulant is localised in latitude, with some recurrent
correlations occurring on the jet spacing, whilst the structure in
longitude contributions both from wavenumber $k_x=1$ and from the
scale of the forcing. Examination of
the spatio-temporal dynamics of the system indicates that the $k_x=1$
contribution arises from a domain-scale meandering of the jet, termed
``satellite modes'' by Ref. \cite{danilovgurarie04}. 
The right panel of Fig.~\ref{figure5} shows that CE2 
reproduces the contributions to the second cumulant at the
longitudinal scales of the forcing, but is incapable of reproducing
the contribution from the satellite modes, when the system is this far
from equilibrium. Interestingly these modes are also absent from
quasilinear DNS calculations \cite{sriyoung12}, which would seem to
indicate that they arise as a result of eddy + eddy $\rightarrow$ eddy
interactions. 

For systems driven even further from equilibrium, CE2 struggles not
only to reproduce all the structures of the second cumulant, but also
the number of jets and their strength. As noted earlier, for smaller
$R_\beta$ the jets 
are more intermittent and meander more. Although zonal averages can be
calculated, the constant meandering of the jets in latitude reduces
the average jet strength. CE2 eventually settles down to a fixed point though we do not believe this to be a unique solution. 
The solution overestimates the strength of
the jets and therefore the Rhines scale associated with them; hence
CE2 has a tendency to underestimate the number of jets as shown in Figs. ~\ref{figure3}(b) and \ref{figure4}(c).

This paper has demonstrated that DSS as approximated by CE2 performs
well in directly calculating the statistics for $\beta$-plane
turbulence in quasi-equilibrium. It confirms the earlier result
\cite{Tobiasetal11,sriyoung12} that zonal jets do not require an
inverse cascade to be driven, but can arise as the result of Reynolds
stresses alone. However, and importantly, we have shown that as the
system is removed further from equilibrium by reducing the zonostrophy
parameter $R_\beta$, CE2 can significantly overestimate jet strengths
and predict the  incorrect number of jets. We hypothesise that for
such systems higher order cumulant expansions are required. If
truncated at third order (CE3) the cumulant expansion includes eddy +
eddy $\rightarrow$ eddy interactions and should perform better in
predicting statistics for out-of equilibrium systems.  The potential
utility of CE3 has been demonstrated for the problems of an isolated
vortex \cite{Chertkov:2010jr} and fluid flow relaxing to a prescribed
jet \cite{Marston:2012co}. We conclude by noting that although we have
stressed the limitations of CE2, we believe that the local
$\beta$-plane system driven stochastically is one of the stiffest
tests of this method; it is very difficult in both DNS and DSS to
reach the quasi-equilibrium limit; although progress may be achieved
utilising DSS implementing semi-implicit timestepping.  Nevertheless
CE2 provides a good qualitative description of the first
  cumulant for systems where the important competing effects arise
from inhomogeneity, anisotropy, and turbulent fluctuations about a
non-trivial basic state.

\begin{acknowledgments}
We wish to acknowledge useful discussion with Nikos Bakas, Freddie Bouchet, David Dritschel, Brian Farrell, Petros Ioannou, Balu Nadiga, 
Wanming Qi, Tapio Schneider, Peter Read, Geoff Vallis, Peter Weichman, and Bill Young.
This work was supported in part by NSF under grant Nos. DMR-0605619 and CCF-1048701 (JBM).   JBM thanks the Aspen Center for Physics 
(supported in part by NSF Grant No. 1066293) for its hospitality during the summer 2012 workshop on  ``Stochastic Flows and Climate Modeling.''
\end{acknowledgments}

%

\end{document}